\documentclass[aps,showpacs,twocolumn]{revtex4}

\usepackage{epsfig}
\usepackage{amsfonts}
\usepackage{amssymb}
\usepackage{mathrsfs}
\usepackage{theorem}
\usepackage{amsmath}
\usepackage{times}
\usepackage{color}

\usepackage{ifpdf}
\ifpdf
\usepackage{epstopdf}   
\fi

\newtheorem{theorem}{Theorem}

\newtheorem{lem}{Lemma}
\newtheorem{corollary}{Corollary}

\newcommand{\ket}[1]{\ensuremath{\vert#1\rangle}}
\newcommand{\bra}[1]{\ensuremath{\langle #1\vert}}
\newcommand{\bk}[2]{\ensuremath{\langle #1\vert #2\rangle}}
\newcommand{\kb}[2]{\ensuremath{\vert #1 \rangle \langle #2 \vert}}

\newcommand{\earl}[1]{#1}  

\newcommand{\etc}[1]{#1}

\def\R{\mathbb{R}}
\def\C{\mathbb{C}}

\renewcommand{\vec}[1]{\ensuremath{\mathbf{#1}}}

\newcommand{\tr}{\ensuremath{\mathrm{tr}}}

\def\id{\mbox{\small 1} \!\! \mbox{1}}

\def\id{\mbox{\small 1} \!\! \mbox{1}}

\def\id{{\mathchoice {\rm 1\mskip-4mu l} {\rm 1\mskip-4mu l} {\rm 1\mskip-4.5mu l} {\rm 1\mskip-5mu l}}}

\begin{document}
\title{Gaussification and Entanglement Distillation of Continuous-Variable Systems:  \\A Unifying Picture}

\author{Earl T.\ Campbell and Jens Eisert}
\affiliation{Dahlem Center for Complex Quantum Systems, Freie Universit{\"a}t Berlin, 14195 Berlin, Germany}
\affiliation{Institute of Physics and Astronomy, University of Potsdam, 14476 Potsdam, Germany}
\pacs{03.67.Bg,42.50.Ex}

\begin{abstract}
Distillation of entanglement using only Gaussian operations is an important primitive in quantum 
communication, quantum repeater architectures, and distributed quantum computing.  
Existing distillation protocols for continuous degrees of freedom are only known to converge 
to a Gaussian state when measurements 
yield precisely the vacuum outcome. In sharp contrast, non-Gaussian states can be 
deterministically converted into Gaussian states while preserving their second moments, 
albeit by usually reducing their degree of entanglement. In this work -- based on a novel instance of a 
non-commutative central limit theorem -- we introduce a picture general
enough to encompass the known protocols leading to Gaussian states, and new classes of protocols including multipartite distillation. This gives the experimental 
option of balancing the merits of success probability against entanglement produced.

\end{abstract}

\maketitle  

Entangled quantum states are the fundamental resources that enable quantum key distribution, quantum communication, and instances of
distributed quantum computing.  Real physical systems are affected by decoherence and non-ideal apparatus that degrades the quality of experimentally preparable quantum states.  However, entanglement distillation protocols provide a means of converting many copies of partially entangled states into a smaller number of more entangled states~\cite{Bennett}.  
When entanglement is required over very long distances, 
distillation can be implemented at regular intervals called repeater nodes~\cite{Repeater,Hybrid,Gisin}.  
Photonic systems that carry entanglement in continuous degrees of freedom are difficult to manipulate arbitrarily.  
However, so-called \textit{Gaussian} operations are more easily implemented by a combination of beam-splitters, phase shifters and squeezers.  Furthermore, preparation of Gaussian states is routine in many laboratories, and such states are especially 
useful for numerous quantum information tasks.  

Unfortunately, Gaussian operations are quite limited in their \earl{capacity to distill entanglement.} 
In particular, a series of no-go theorems have shown that with only
Gaussian resources, it is impossible to \earl{increase entanglement}~\cite{Nogo}.  These results can be circumvented \earl{when non-Gaussian} resources are available.  Given an appropriate resource, Gaussian operations can 
simultaneously increase the entanglement and make the state more Gaussian, 
a process we refer to as the {\it ``Gaussification protocol''}
(GP) \cite{Gaussification} (for steps towards experimental realization of this and related protocols, see Refs.\ \cite{Exp}).  Alternatively, the theorem can also be circumvented by using a non-Gaussian operation implemented by photon detectors.  Entanglement distillation can be achieved by a combination of photon subtraction, a de-Gaussifying operation, and Gaussification~\cite{Gaussification,Fiur10}.

While these techniques give hope for simple realizations of quantum information protocols, they do not exploit the richness of Gaussian operations.   Specifically, GP utilizes only projections onto the vacuum.  These projections are feasible if reliable detectors are available that distinguish zero from one or more photons.  However, strictly speaking these detectors do not fall within the realm of Gaussian devices.  Performing eight-port homodyne detection~\cite{Leonhardt} and postselecting on the vacuum outcome achieves the same projection, but this will have zero success probability when postselecting \textit{exactly} on the vacuum measurement outcomes.  In this work, we prove Gaussification for a wide class of truly Gaussian protocols with nonzero, and tunable, success probabilities for multimode states.  \etc{All known feasible distillation protocols, including our protocols, are iterative and consume a number of copies exponential in the iterations required.  As such, our scheme's capability to increase success probability significantly improves the prospects of distillation and repeater implementations with only modest resources.  Our techniques also open} up the perspective of directly distilling into multi-partite Gaussian states.

% All known practical distillation protocols, including our protocols, are iterative, consume an exponential number of copies per round and have an overall success that is quite sensitive to the success probability per round.  As such, capability of our scheme to increase success probability is highly desirable and makes the prospect of practical implementations more promising

 \begin{figure}
 \includegraphics[width=7.2cm]{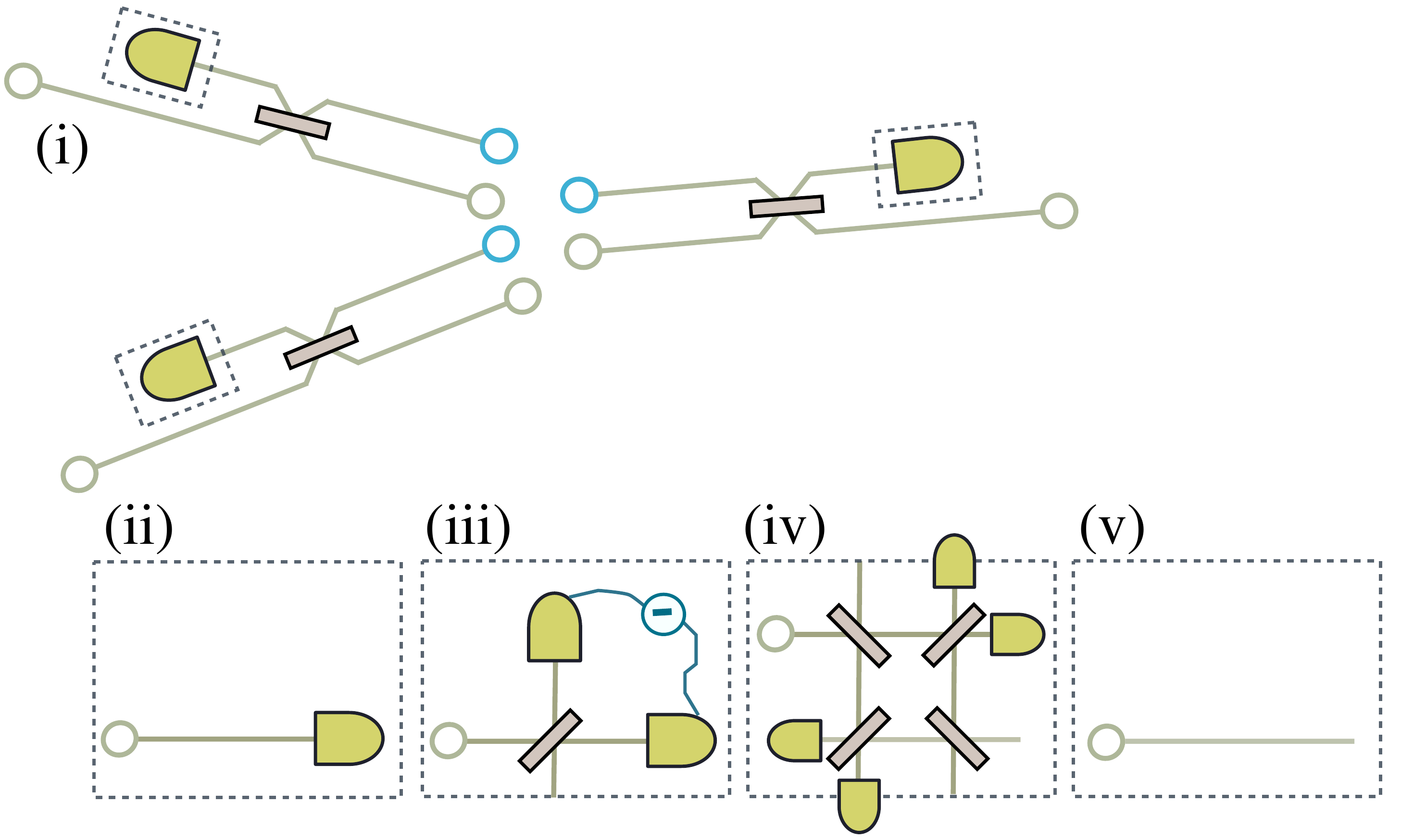}
 \caption{(i) A single step of the general class of protocols (GG) considered, illustrated for 
 three parties. This embodies the known Gaussifier (GP) entanglement distillation schemes based on
 projections onto the vacuum, or the extremality protocol (EP) mapping unknown states onto Gaussian ones with the same second moments.
 The Gaussian projection can be arbitrary (including (ii) vacuum projection, (iii) homodyning, (iv) eight-port homodyning, (v) tracing out); and schemes with finite widths of acceptance.}
 \label{fig:overview}
 \end{figure}

% In addition to the practical applications of our results, we also provide some deep insights into why Gaussian states emerge in quantum
%information protocols.  

In addition to the practical applications of our results, \etc{our analysis provides a more intuitive explanation of the phenomena of Gaussification.  In the original distillation protocols~\cite{Gaussification}, the process of Gaussification is quite mysterious,} but it is \earl{very apparent} in the \earl{protocol} of Ref.\ \cite{Wolf} which provides an alternative method that uses no measurements, referred to as \earl{the} {\it ``extremality protocol''} (EP), as it is used to show the extremality of Gaussian states with respect to several properties.  \earl{Yet} EP can convert many non-Gaussian states into a more Gaussian state while conserving the expectation value of observables quadratic in position and momentum.  Although the EP does Gaussify, its capacity for increasing entanglement is restrained by its deterministic nature. The proof of EP elegantly employs the central limit theorem~\cite{CLT} that explains the ubiquity of Gaussian distributions in classical statistics and nature itself.  Our approach unifies GP and EP within a comprehensive theory of {\it ``general Gaussification''} (GG) founded on a non-commutative central limit theorem \etc{and so provides an intuitive mechanism for Gaussification.} In addition to bipartite entanglement distillation, our approach reveals whole new classes of protocols that we discuss.

We consider GG protocols that can be implemented iteratively, with the $(n+1)^{\mathrm{
th}}$ iteration as follows:  
\begin{enumerate}
  \setlength{\itemsep}{0pt}
  \setlength{\parskip}{0pt}
  \setlength{\parsep}{0pt}
\item Take two copies of an $m$-mode state $\rho_{n}$ shared between $m$ parties; 
\item Each party applies a 50:50 beam-splitter transformation between their pair of modes; 
\item Every party makes a Gaussian measurement on the output of one beam-splitter port; 
\item The parties compare measurement results and postselect such that the operation implemented is Gaussian; 
\item The output state $\rho_{n+1}$ is used for the next iteration.
\end{enumerate}

{\it Formal description of GG protocols.}
There are $2m$ modes involved in the protocol, and we label annihilation operators,  $ \hat{a}_{j,k} $, with two indices;   The index $j=1,2,\dots,m$ labels the respective party and $k=1,2$ the copy at a particular node.  The beam-splitters, in the Heisenberg picture, perform
\begin{eqnarray}
\label{eqn:beamsplitt}
	U  \hat{a}_{j,k} U^{\dagger} = ( \hat{a}_{j,1} +(-1)^{k} \hat{a}_{j,2})/\sqrt{2}.
\end{eqnarray}
The measurements at step 3 can be homodyne, eight-port homodyne, or any other Gaussian measurements 
projecting onto a state $\Pi_{\vec{m}}$ for measurement outcome $\vec{m}$.  We are interested in Gaussian protocols 
that postselect on a set of measurement outcomes, and mix over all accepting outcomes, 
\begin{equation}
	\rho_{n+1} \propto \int d\vec{m} P( \vec{m} ) \tr^{2} [ U (\rho_{n} \otimes \rho_{n} ) U^{\dagger} (\id \otimes \Pi_{\vec{m}})  ] 
\end{equation}
where $\tr^{2}$ denotes a partial trace over the second copy, with $k=2$.   The integral over measurement outcomes is weighted by  $P( \vec{m} )$.  The weights $P( \vec{m} )=0$ and $P( \vec{m} )=1$ correspond to a rejection and an acceptance, respectively, but we also allow for probabilistic strategies where $P( \vec{m} )$ gives the probability of acceptance.  The protocol is described by a single operator we call the {\it filter}
\begin{equation} 
	\Pi = \int d\vec{m}  P(\vec{m}) \Pi_{\vec{m}} .
\end{equation}
We allow for arbitrary Gaussian filters, $\Pi$, that are invertible and proportional to 
a fully separable Gaussian state with vanishing first moments and finite energy. 
Certain interesting cases, such as GP, EP and protocols using precise homodyne detection are included as limits within this family of filters, so keeping full generality. 
With this notation
\begin{equation}
\label{eq:iterate}
	\rho_{n+1} \propto  \tr^{2} [ U (\rho_{n} \otimes \rho_{n}) U^{\dagger} (\id \otimes \Pi)  ] .
\end{equation}

{\it Phase space.}
Before we give our results, we review phase space representations for the position and momentum observables of an $m$-mode system, labeled as
\begin{equation}
	\hat{\vec{R}}=( \hat{R}_{1}, \hat{R}_{2},\dots , 
	\hat{R}_{2m-1}, \hat{R}_{2m}  )= ( \hat{X}_{1}, \hat{P}_{1},\dots, \hat{X}_{m}, \hat{P}_{m}  ), \nonumber
\end{equation}
where $\hat X_j = (\hat a_j^\dagger + \hat a_j)/\sqrt{2}$ and $\hat P_j = i (\hat a_j^\dagger - \hat a_j)/\sqrt{2}$ for $j=1,\dots, m$. The canonical commutation relations between the coordinates
are embodied in the symplectic matrix $\Sigma$.
The {\it covariance matrix} of an operator $A$ records the second moments of these observables.
\begin{equation}
	(\Gamma_{A} )_{j,k}= \tr (  \{ \hat{R}_{j}-(\vec{d}_{A})_{j} , \hat{R}_{k}-(\vec{d}_{A})_{k} \}_{+} A ) ,
\end{equation}
where $\{. , .\}_{+}$ denotes the anticommutator and first moments are $(\vec{d}_{A})_{j} = \tr ( \hat{R}_{j} A )$.  Furthermore, we make use of characteristic functions, $\chi_{A}: \R^{2m}\rightarrow\C$, that encode all the information of $A$ as $\chi_{A}(\vec{r}) = \tr ( D(\vec{r}) A )$, where $D(\vec{r})$ is the displacement operator, $D(\vec{r})=\exp ( i \vec{r}\cdot  \hat{\vec{R}} )$.   Such a function is said to be Gaussian when
\begin{equation}
	\chi_{A}(\vec{r}) = \exp ( i \vec{r}\cdot \vec{d}_{A} - \vec{r}^{T} \Gamma_{A} \vec{r}/4 ) .
\end{equation}
If the operator $A$ is a physical state, its covariance matrix will be real.  However, an instrumental tool in our analysis is that we work with $A=\sigma_{n}=\rho_{n}\Pi/\tr(\rho_{n}\Pi)$. Indeed,
we will not employ characteristic functions of states satisfying the conditions of Bochner's theorem \cite{CLT,Davies}, but of more general objects, hence leading  to more general complex-valued functions.
Since $\Pi$ 
is invertible, $\sigma_{n}$ uniquely defines a quantum state $\rho_{n}$.  

{\it A new non-commutative quantum central limit theorem.}  
With these definitions at hand we can state our first result, with a stronger form of convergence demonstrated later.
\begin{theorem}[Convergence of general Gaussifier protocols]
\label{thm1}
Consider an initial state $\rho=\rho_{0}$, with associated operator $\sigma=\rho \Pi / \tr( \rho \Pi)$ such that the following conditions are satisfied: (i) $\vec{d}_{\sigma}=0$; (ii) $| \chi_{\sigma}(\vec{r}) | \leq 1$ for all $\vec{r}$; (iii) the covariance matrix
 \begin{equation}
\label{eqn:convergent_gamma}
	\Gamma_{\rho_{\infty}} =  (  \Gamma_\Pi -  i \Sigma )(  \Gamma_\Pi -\Gamma _{\sigma})^{-1}(  \Gamma_\Pi + i \Sigma ) - \Gamma_\Pi ,
\end{equation}
exists and is positive definite.
Let $\rho_{\infty}$ denote the Gaussian state with covariance matrix $\Gamma_{\rho_{\infty}}$.  GG with filter $\Pi$ causes $\rho_{n}$ to weakly converge to $\rho_{\infty}$ in the following sense:  If $\ket{x}$ and $\ket{y}$ are eigenvectors of $\Pi$, then for all $\epsilon>0$ there exists a $n_{\epsilon}$ such that for all $n>n_{\epsilon}$
\begin{equation}
	| \bra{x} \rho_{n}\ket{y}/ \tr( \rho_{n} \Pi) - \bra{x} \rho_{\infty}\ket{y}/ 
	\tr( \rho_{\infty} \Pi) | < \epsilon.
\end{equation}
\end{theorem}
{\it Proof of the statement.} 
Much of the basic structure of the proof follows an argument of a non-commutative 
quantum central limit theorem for quantum states \cite{CLT,Davies,Wolf,MCLT}.
The problem once one allows for Gaussian measurements is that the characteristic function of the output is unwieldy indeed. We circumvent this problem by a bold step: We put in an additional filter $\Pi$ at the output ``by hand'', in order to exploit symmetry, at the expense of having to consider $\chi_{\sigma_n}$ of different objects, $\sigma_{n}$. \earl{This will then lead to the desired result.  From Eq.~(\ref{eq:iterate}) we have,
\begin{equation}
	\chi_{\sigma_{n+1}}(\vec{r}) \propto  \tr [ ( D(\vec{r}) \otimes \id )   U ( \rho_{n} \otimes \rho_{n}) U^{\dagger} (\Pi \otimes \Pi) ] .
\end{equation}}
Using the cyclicity of the trace and Eq.~(\ref{eqn:beamsplitt})
\begin{equation}
	\chi_{\sigma_{n+1}}(\vec{r}) \propto  \tr [  D(\vec{r}/\sqrt{2})^{\otimes 2} \rho_{n}^{\otimes 2}U^{\dagger} \Pi^{\otimes 2} U  ] .
\end{equation}
Next we recall that Gaussian states with zero 
first moments commute with beam-splitters, $U \Pi^{\otimes 2}  = \Pi^{\otimes 2} U$, such that
\begin{eqnarray}
	\chi_{\sigma_{n+1}}(\vec{r}) & \propto &  \tr \{ [ D(\vec{r}/\sqrt{2}) \rho_{n}\Pi  ]^{\otimes 2} \}   \propto   \tr [  D(\vec{r}/\sqrt{2}) \sigma_{n}]^{2} , \nonumber \\ 
	& = &  \chi_{\sigma_{n}}  (  \vec{r} / \sqrt{2}  )^{2}  =  \chi_{\sigma}  (  \vec{r} / \sqrt{N}  
	)^{N} ,
\end{eqnarray}
where in the last equality $N=2^{n+1}$, iterating the formula. 
The key to the simplicity of this formula is to consider convergence of $\sigma_{n}$ rather than directly $\rho_{n}$.  By introducing an additional projector within the trace, 
symmetry allows us to commute through the beam-splitter unitaries, which is the essential simplifying step.  To find the limiting characteristic function, we consider a given phase space point $\vec{r}_{0}$ and the function
$f_{\sigma_{n+1},\vec{r}_{0}}:\R\rightarrow \C$ defined as $f_{\sigma_{n+1},\vec{r}_{0}}(t)= \chi_{\sigma_{n+1}}(t \vec{r}_{0} )$.  In the spirit of a {\it classical central limit theorem} \cite{CLT,Wolf,MCLT} but for non-Hermitian operators we write
\begin{eqnarray}
	f_{\sigma,\vec{r}_{0}}\left( \frac{t}{\sqrt{N}} \right)^{N} & \propto &  \left( 1 -t^{2} 
	\frac{ \vec{r}_{0}^{T} \Gamma_{\sigma} \vec{r}_{0}}{4 N} + 
	o\left( \frac{t^{2}}{N} \right)  \right) ^{N} ,
\end{eqnarray}
which converges assuming that second moments are finite, $f(0)=1$ and $| f(t) | \leq 1$ for all $t$.  This last condition, which is always satisfied for classical characteristic functions, may be violated for non-Hermitian 
$\sigma_{n}$.  However, provided $| \chi_{\sigma} (\vec{r}) | \leq 1$ for all $\vec{r}$,  we find (see App.~\ref{APP:CLT}) in the limit of large $n$ that 
\begin{eqnarray}
\label{eqn:converges_too}
	\lim_{n \rightarrow \infty} f_{\sigma_{n+1}}(t) & \propto & \exp ( - t^{2} \vec{r}_{0}^{T} \Gamma_{\sigma} \vec{r}_{0} / 4 ),
\end{eqnarray}
pointwise in $t$.  Setting $t=1$ shows pointwise convergence of $\chi_{\sigma_{n+1}}$ for each phase space point $\vec{r}_0$.  Furthermore, following the reasoning of Refs.\ \cite{Davies,MCLT}, this entails that any operator $B$ that is absolutely integrable such that $\int | \chi_{B} (\vec{r})| d \mathbf{r}  < \infty$, has a convergent expectation value $ \tr(B \sigma_{n}) \rightarrow \tr(B \sigma_{\infty})$  (see App.~\ref{APP:expectationvalues}).  Setting $B=\lambda_{y}^{-1} \kb{y}{x}$ where $\Pi \ket{y}=\lambda_{y} \ket{y}$, then for large $n$
\begin{equation}
	  \frac{\bra{x} \rho_{n} \ket{y}}{\tr( \rho_{n} \Pi ) } =  \frac{\bra{x} \sigma_{n} \ket{y}}{\lambda_{y}} \rightarrow  \frac{\bra{x} \sigma_{\infty} \ket{y}}{\lambda_{y}} =  \frac{\bra{x} \rho_{\infty} \ket{y}}{\tr( \rho_{\infty} \Pi ) }  .
\end{equation}
All that remains is to show that Gaussian $ \sigma_{\infty} $ entails a Gaussian $\rho_{\infty}$. First we observe (see App.~\ref{APP:GaussANDSchur}) that 
the product of two operators has a characteristic equation 
\begin{equation}
\label{eqn:fourier_convolve}
 	\chi_{\sigma_{\infty}} (\vec{q}) \propto \int \chi_{\rho_{\infty}} (\vec{r}) \chi_{\Pi} (\vec{q}-\vec{r}) \exp ( -i \vec{r}^{T} \sigma \vec{q}/2 ) d\vec{r} .
\end{equation}
For Gaussian $\chi_{\Pi}$ and $\chi_{\rho_{\infty}}$, the integral is a multi-variate Gaussian integral that evaluates (see App.~\ref{APP:GaussANDSchur}) to 
another Gaussian with the covariance matrix taking the form of a Schur complement~\cite{Horn},
\begin{equation}
\label{eqn:Schur} 
	\Gamma_{\sigma} = 
	\Gamma_\Pi - (\Gamma_\Pi + i \Sigma )(\Gamma_{\rho_{\infty}} 
	+ \Gamma_\Pi)^{-1}(\Gamma_\Pi-i \Sigma ) ,
\end{equation}
Rearranging this formula for $\Gamma_{\rho_{\infty}}$ gives us the covariance matrix for the convergent state $\rho_{\infty}$ in terms of $\Gamma_{\sigma}$ and $\Gamma_\Pi$ as in the theorem.

\begin{figure}
\centering
\includegraphics[width=8cm]{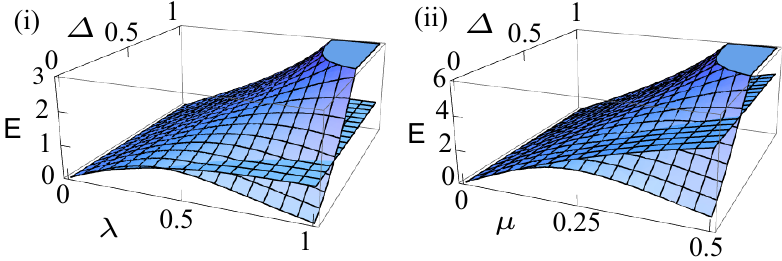}
\caption{The degree of entanglement of the initial state $\rho$ in terms of the log-negativity
versus the target Gaussian $\rho_{\infty}$ for 
(i)  bipartite state vectors $\ket{\Psi_{\lambda}}$ and (ii) tripartite state vectors $\ket{\Phi_{\mu}}$.  In both figures the variable $0 < \Delta < 1$ controls the degree of postselection.  The curve varying with $\Delta$ is the entanglement of $\rho_{\infty}$, and the curve constant in $\Delta$ is the initial entanglement of $\rho$. }
\label{fig:3plots}
\end{figure}
  
{\it  Examples.} 
We have introduced a broad class of protocols for which our theorem indicates Gaussification.  However, for concreteness it is helpful to keep in mind a simple class of protocols.  Consider when each party performs eight-port homodyne measurements that project onto a coherent state.  When the $k^{\mathrm{th}}$ party obtains outcome $\alpha_{k}$ projecting onto coherent state vector $\ket{\alpha_{k}}$, we declare the iteration a success with probability $P=\exp(-| \boldsymbol{ \alpha} |^{2}/2 c^{2})$.  The degree of postselection is quantified by a real variance $c$.  It follows that the filter is indeed proportional to a Gaussian state with covariance matrix $\Gamma_\Pi = \Delta^{-1} \id $, with $\Delta=(1+c^{2})^{3/2}$.  This class of protocols is important as it contains GP and EP as limits $\Delta \rightarrow 1$ and $\Delta \rightarrow 0$, respectively. 

We now consider the degree of entanglement that is achieved by applying our 
protocol with eight-port homodyne measurements.  First we consider the well-studied bipartite state vector $\ket{\Psi_{\lambda}} \propto \ket{0,0}+\lambda \ket{1,1}$, and present the log-negativity~\cite{neg} of the convergent Gaussian in Fig.~\ref{fig:3plots}.  Varying the parameter $\Delta$ interpolates between the entanglement achieved by GP and EP, with increased yield compensating for reductions in entanglement.   We also analyzed a tripartite entangled state  vector $\ket{\Phi_{\mu}}\propto \ket{0,0,0}+\mu(\ket{1,1,0}+\ket{1,0,1}+\ket{0,1,1})$, having up to 
two photons in three modes, and for the log-negativity summed over all $3$ bipartitions. In fact, in this way, one can straightforwardly engineer
multi-partite hybrid distillation protocols for quantum networks, giving rise to primitives in
 repeater architectures where entanglement is shared across many repeater nodes.  
These would overcome the known limitations providing a road-block 
against entirely {\it Gaussian continuous-variable repeater 
networks} \cite{Ohliger}. 

{\it Strong convergence.} 
Our previous theorem proves a convergence result identical to that of \earl{Ref.~\cite{Gaussification}} and GP.   
However, it would often be preferable to have convergence of $\bra{x} \rho_{n}\ket{y}$ to $\bra{x} \rho_{\infty}\ket{y}$ in a stronger sense (without the addition factors of $\tr( \rho_{n}\Pi)^{-1}$ and $\tr( \rho_{\infty}\Pi)^{-1}$ and as a convergence in trace-norm). For most physically
relevant (see App.~\ref{APP:weaktostrong}) instances of initial quantum states, we now show that this is indeed the case. We will make use of expectation values of normally order operators,
\begin{equation}
	\alpha_{n}^{\vec{x}, \vec{y}} = \tr [ V ( \otimes_{k=1}^{m} \hat{a}_{k}^{x_{k}} )^{\dagger} ( \otimes_{j=1}^{m} \hat{a}_{j}^{y_{j}} ) V^{\dagger}  \sigma_{n} ],
\end{equation}
where $\vec{x}, \vec{y} \in \mathbb{N}^{m} $ and $V$ is the Gaussian unitary such that $V \Pi V^{\dagger}$ is a thermal state.

\begin{theorem}[Strong convergence]
\label{thm2}
In addition to Theorem \ref{thm1}, if for all $\vec{x}, \vec{y} \in \mathbb{N}^{m} $ the expectations values of normally ordered operators satisfy,  $| \alpha_{0}^{\vec{x}, \vec{y}} |   \leq \alpha_{\infty}^{\vec{x}, \vec{y}}  $
for 
$\alpha_{\infty}^{\vec{x}, \vec{y}}   = |\alpha_{\infty}^{\vec{x}, \vec{y}} |$. It follows that for all $\epsilon>0$ there exists an $n_{\epsilon}$ such that for all $n>n_{\epsilon}$,  we have $ \| \rho_{n} - \rho_{\infty}  \| < \epsilon$, where $ \| .  \|$ is the trace norm.
\end{theorem}
The conditions of the theorem are stated technically, but physically prevent overpopulation of higher Fock numbers.  E.g., it is easy to check these conditions are meet in all the low photon examples analyzed in Fig~\ref{fig:3plots}. We begin by first showing $\tr(\Pi \rho_{n})$ converges to $\tr(\Pi \rho_{\infty})$, which in terms of $\sigma$ is
\begin{equation}
	\tr ( \Pi^{-1} \sigma) = \tr ( \Pi^{-1}  \rho \Pi) / \tr ( \rho \Pi) = \tr ( \rho \Pi)^{-1} .
\end{equation} 
The inverse filter, $\Pi^{-1}$, has an exponential form that can be Taylor expanded and, using the bosonic commutation relation, normally ordered such that $\tr ( \Pi^{-1} \sigma_{n} ) = \sum_{\vec{x}, \vec{y}} q_{\vec{x}, \vec{y}}  \alpha_{n}^{\vec{x}, \vec{y}}$,
where $q_{\vec{x}, \vec{y}}\geq 0$.  Hence, when the conditions of our theorem hold, we conclude $\tr( \Pi^{-1} \sigma_{0}) \leq \tr( \Pi^{-1} \sigma_{\infty})$, and we proceed by showing this holds for all $\sigma_{n}$.  Iteratively we have (details in App.~\ref{AppIt})
\begin{eqnarray}
\label{EqIt}
	\alpha_{n+1}^{\vec{x}, \vec{y}} =  \sum_{\vec{u}\leq \vec{x}, \vec{v} \leq \vec{y}} C_{\vec{u}, \vec{v}}^{\vec{x}, \vec{y}} \alpha_{n}^{\vec{u}, \vec{v}} \alpha_{n}^{\vec{x-u}, \vec{y-v}} ,
\end{eqnarray}
where $C_{\vec{u}, \vec{v}}^{\vec{x}, \vec{y}} $ is a combinatorial quantity that is non-negative and real.  Under our assumptions the absolute values obey
\begin{eqnarray}
	|\alpha_{n+1}^{\vec{x}, \vec{y}}| \leq  \sum_{\vec{u}\leq \vec{x}, \vec{v} \leq \vec{y}} C_{\vec{u}, 
	\vec{v}}^{\vec{x}, \vec{y}} \alpha_{\infty}^{\vec{u}, \vec{v}} 
	\alpha_{\infty}^{\vec{x-u}, \vec{y-v}} = \alpha_{\infty}^{\vec{x}, \vec{y}},
\end{eqnarray}
so initially satisfying our conditions entails satisfaction for all $n$. Hence, for all $n$ we deduce $\tr( \Pi^{-1} \sigma_{n}) \leq \tr( \Pi^{-1} \sigma_{\infty})  $ and equivalently  $\tr ( \rho_{n} \Pi ) \geq \tr ( \rho_{\infty} \Pi )  $.  
Next we bound $\tr ( \rho_{n} \Pi ) $ from above. Consider a finite rank projector $P$ that commutes with $\Pi$, then $B=\Pi^{-1}P$ has an absolutely integrable characteristic function. Hence, for arbitrarily small $\delta>0$ there exists an $n_{\delta}$ such that for $n>n_{\delta}$
\begin{equation}
	\tr(B\sigma_{n})-\tr(B\sigma_{\infty}) = \frac{\tr( P\rho_{n} ) }{\tr( \Pi \rho_{n})}- \frac{\tr( P\rho_{\infty} ) }{\tr( \Pi \rho_{\infty})} \geq - \delta .
\end{equation}
Since $\tr( P \rho_{n}) \leq 1$, and $P$ can be chosen so $\tr( P\rho_{\infty})=1-\delta' $ is arbitrarily close to unity, we conclude $\tr ( \Pi \rho_{n} )^{-1}-\tr(\Pi \rho_{\infty})^{-1} \geq - \epsilon$ where $\epsilon= \delta+\delta' \tr(\rho_{\infty} \Pi)^{-1}$ is again small.   This inequality rearranges to $\tr ( \Pi \rho_{n}) \leq ( \tr ( \Pi \rho_{\infty} )^{-1} - \epsilon)^{-1} $, giving an arbitrarily tight bound from above.  Combined with our lower bound we conclude that $\tr ( \Pi \rho_{n})$, under the stated assumptions, converges to $\tr ( \Pi \rho_{\infty}) $.  As such $\bra{x} \rho_{n} \ket{y}$ converges to $\bra{x} \rho_{\infty} \ket{y}$, and as is well known this entails trace norm convergence (See Ref.~\cite{Davies} and App~\ref{TraceNorm}). 

{\it Summary.} 
We have introduced a framework for constructing a range of new protocols for entanglement distillation and manipulation. At the same time, this work provides
a unified framework for existing protocols leading to Gaussian states: Notably, the mysterious emergence of
Gaussian states in distillation schemes is once again related to an instance of a quantum 
central limit theorem, albeit for a much more broader class of protocols than previously considered. This framework also allows us to 
look at Gaussification in experimentally realistic acceptance windows, and to trade-off different figures of merit
against each other. Such trade-off control is essential as previous proposals are so heavily postselective that over several iterations the success probability would reduce dramatically, whereas our protocols offer an arbitrarily good chance success. Potential for future research is broad as a unique protocol is defined by every separable Gaussian state.  For example, our techniques can be applied to homodyne detection protocols for either multimode entanglement distillation, or single mode squeezing enhancement.

{\it Acknowledgements.} 
We would like to thank the EU (QESSENCE, MINOS, COMPAS), the BMBF (QuOReP), and the EURYI for support, and thank A.\ Mari, C.\ Gogolin, V.\ Nesme, \earl{D.\ E.\ Browne} 
and M.\ Ohliger for valuable discussions.

\appendix

\section{Central limit theorems reviewed}
\label{APP:CLT}

Here we present additional details on the convergence of characteristic functions to a Gaussian function.  Our proof follows the classical proofs, such as in Ref.~\cite{Moran}, but since $\sigma = \rho \Pi / \tr( \rho \Pi )$ is generally neither positive nor Hermitian there are some notable subtleties and technicalities
involved. Ultimately, we aim to prove the following. 
\begin{theorem}[General quantum central theorem]
\label{thm_QCLT}
	If $f_{\sigma,\vec{{r}_0}}(t)$ satisfies $| f_{\sigma,\vec{{r}_0}}(t) | \leq 1$ for all 
	$t\in \R$ then in the limit of large $N$
	\begin{equation}
		 f_{\sigma,\vec{{r}_0}}(t/\sqrt{N})^{N}  \rightarrow e^{-\nu t^{2}/4} ,
	\end{equation}
	pointwise in $t$, with $\nu$ as the appropriate second moment.
	Specifically, pointwise convergence means that for all $t\in \R$ 
	 and all $\epsilon>0$ there exists a $N_{\epsilon}$ such that for all $N>N_{\epsilon}$ we have
	 \begin{equation}	
	 	| f_{\sigma,\vec{{r}_0}}(t/\sqrt{N})^{N} - e^{-\nu t^{2}/4} | \leq \epsilon .
	 \end{equation}
\end{theorem}
Setting $t=1$ directly entails pointwise convergence of the characteristic function $\chi_{\sigma_{n}}$, and furthermore this is uniform on compact regions, such as a ball in phase space. 
\begin{corollary}[Convergence in compact regions]
\label{cor_uniform_converge}
	Within any compact region $R$, $\chi_{\sigma_{n}}$ 
	converges uniformly to  $\chi_{\sigma_{\infty}}$. 
	That is, for any $\epsilon>0$ 
	there exists an $n_{\epsilon}$ such that for all 
	$n>n_{\epsilon}$  and all $\vec{r} \in R$, 
	we have $|  \chi_{\sigma_{n}} (\vec{r}) - \chi_{\sigma_{\infty}}(\vec{r})| \leq \epsilon$ .
\end{corollary}
Pointwise convergence is uniform on compact regions whenever the gradient is bounded 
within $R$ for all $n$.  To prove a gradient bound for all $n$, we first observe that for $n=0$ a bounded continuos function always has a gradient bound on compact sets.  That is, an $\eta$ can always be found such that for all $\vec{r} \in R$ we have
$G_{0}(\vec{r}) =\|\nabla \chi_{\sigma_{0}}(\vec{r})\|_2< \eta  |\vec{r}|$, where $\|.\|_2$ denotes the 2-norm.   This property is iteratively preserved for all $\sigma_{n}$, since
\begin{eqnarray*}
	G_{n}(\vec{r}) & = &  \|\nabla \chi_{\sigma_{0}}(\vec{r}/\sqrt{N})^{N}  \|_2 ,  \\ \nonumber
	& = & N \| [ \nabla \chi_{\sigma_{0}}(\vec{r}/\sqrt{N}) ].[ \chi_{\sigma_{0}}(\vec{r}/\sqrt{N})^{N-1}] \|_2 , \\ \nonumber
	& \leq & \sqrt{N} G_{0}(\vec{r}/\sqrt{N}) \leq \eta  |\vec{r}| .
\end{eqnarray*}
Using the point in the compact region that gives the largest value $|\vec{r}|$ giving the desired gradient bound for all $\vec{r} \in R$Ê and all $n$.  This corollary will prove useful in App. \ref{APP:expectationvalues}.

We prove several useful lemmas before directly addressing this proof.  First we observe,
\begin{eqnarray}
\label{def_little_f}
	f_{\sigma,\vec{{r}_0}}(t) & = & \tr [ D( t \vec{r}_{0}) \sigma  ]  , \\
	& = & \int_{- \infty}^{\infty} dx e^{ i x t} F_{\sigma, \vec{r}_0} ( x ),\nonumber
\end{eqnarray} 
where $F_{\sigma, \vec{r}_0} ( x ) $ is the representation of $\sigma$ along a line in phase space
defined by the unit vector ${\vec{r}_0} $ (e.g., the position or momentum representation in case
of a unit vector pointing along the axis of phase space).  This is similar to a probability density for measurement of a particular quadrature;  
however,  because $\sigma$ is neither positive nor Hermitian, 
the function $F_{\sigma,\vec{r}_0}$ is not generally positive or real, 
which is the main cause for caution.  Yet it is sufficiently well behaved, with the two following lemmas proving useful.

\begin{lem}[Convergence of absolute integrals]
\label{lem_absint}
The absolute integral satisfies $\int_{- \infty}^{\infty}  x^{2} | F_{\sigma, \vec{r}_{0}}(x) | dx<\infty$.
\end{lem}
Now clearly
$(\Pi-i \rho)(\Pi+i \rho)\geq 0$,
such that
\begin{equation}
	\Pi^2 + \rho^2 \geq i\rho \Pi - i\Pi \rho 
\end{equation}	
and 
$(\Pi - \rho)(\Pi -\rho)\geq 0$,
so that 
\begin{equation}
	\Pi^2 + \rho^2 \geq \Pi \rho + \rho \Pi.
\end{equation}	
Similarly, $\Pi^2 + \rho^2 \geq -i\rho \Pi + i\Pi \rho $ and
$\Pi^2 + \rho^2 \geq - \Pi \rho - \rho \Pi$.  One finds after a few steps that, for any vector 
$\ket{\psi}$, we have
\begin{equation}
	| \bra{\psi} \rho \Pi \ket{\psi } | \leq  | \bra{\psi} (\rho^{2}+ \Pi^{2}) \ket{\psi }   | / \sqrt{2}.
\end{equation}
Hence, using the triangle inequality, 
\begin{equation}
	|F_{\sigma,\vec{{r}_0}}(x)|\leq \frac{  | F_{\Pi^{2}, \vec{r}_0}(x) | + | F_{\rho^{2}, \vec{r}_0}(x) | }{\sqrt{2} |\tr ( \rho \Pi )|}.
\end{equation}
That is to say, whenever the second moments of $\Pi^{2}$ and $\rho^{2}$ are finite,
the absolute integral converges to a finite value.

The second lemma we require is as follows.
\begin{lem}[Expansion of functions]
For a given unit vector $\vec{{r}_0}$ and an operator $\sigma$ that is unit trace and zero displacement $\vec{d}_{\sigma}=0$, the above function $f_{\sigma,\vec{{r}_0}}$ can be expressed as
\begin{equation}
	f_{\sigma,\vec{{r}_0}}(t) = 1 -  \frac{1}{4} \nu t^{2} K(t)
\end{equation}
where $K(t) \rightarrow 1$ as $t \rightarrow 0$.  
\end{lem}

This lemma asserts that higher order moments drop off sufficiently quickly for the purposes of establishing a central limit theorem.   We make use of the following
formulation of the exponential function
\begin{equation}
\label{eqn:exp_expand}
	e^{i t x } = 1+ itx - \frac{1}{2} t^{2}x^{2} L(x,t)
\end{equation}
where 
\begin{equation}
	L(x,t) = 2 \int_{0}^{1} (1-u) e^{iutx} du,
\end{equation}
satisfying 
\begin{equation}
	|L(x,t)| \leq 1 
\end{equation}
for $x,t \in \R$. 
Furthermore, for any finite $x$ interval, $e^{iutx} \rightarrow 1$ as $t \rightarrow 0$ uniformly in $x$.  
Consequently, we also have $L(x,t) \rightarrow 1$ as $t \rightarrow 0$ uniformly on any finite $x$-interval.  
Multiplying Eq.~($\ref{eqn:exp_expand}$) by $F_{\sigma,\vec{{r}_0}}$ and integrating w.r.t.\ $x$ and using the unit trace of $\sigma$ and vanishing first moments we have
\begin{equation}
	f_{\sigma,\vec{{r}_0}}(t) = 1 - \frac{1}{2} t^{2} \int_{- \infty}^{\infty}  
	x^{2} L(x,t) F_{\sigma,\vec{{r}_0}}(x) dx ,
\end{equation}
which we can rewrite as
\begin{equation}
	f_{\sigma,\vec{{r}_0}}(t) = 1 - \frac{1}{2} t^{2} \nu K(t) dx ,
\end{equation}
where $\nu$ is the second moment defined by
\begin{equation}
\label{eqn_second_moment}
	\nu = 2 \int_{- \infty}^{\infty} x^{2} F_{\sigma,\vec{{r}_0}}(x) ,
\end{equation}
and $K( t )$ is the quantity satisfying 
\begin{equation}
	\nu K( t ) =  2 \int_{- \infty}^{\infty}  x^{2} L(x,t) F_{\sigma,\vec{{r}_0}}(x) .
\end{equation}
If we subtract Eq.~(\ref{eqn_second_moment}) from the above, we have
\begin{equation}
	\nu( K( t ) - 1 ) = 2  \int_{- \infty}^{\infty}  x^{2} (L(x,t)-1) F_{\sigma,\vec{{r}_0}}(x) .
\end{equation}
The absolute value satisfies
\begin{equation}
	\nu | K( t ) - 1 | \leq 2  \int_{- \infty}^{\infty}  x^{2} | (L(x,t)-1) F_{\sigma,\vec{{r}_0}}(x)| .
\end{equation}
Noting that $|L(x,t)-1| \leq 2$ since $|L(x,t)| \leq 1$, and dividing the integral into two parts over some finite interval $R$ and the complement $R^{\perp}$ we deduce that
\begin{eqnarray}
	\nu | K( t ) - 1 | & \leq & 2 \int_{R}  x^{2} | (L(x,t)-1) F_{\sigma,\vec{{r}_0}}(x)| \nonumber \\
	&  +& 4 \int_{R^{\perp}}  x^{2} |  F_{\sigma,\vec{{r}_0}}(x)|.
\end{eqnarray}
We now make use of Lem.~(\ref{lem_absint}) which tells us that the second integral always converges to a finite value.  Therefore, for any $\epsilon>0$ the region $R$ can be choose sufficiently large that the second integral is bounded such that
\begin{eqnarray}
	\nu | K( t ) - 1 | & \leq & 2 \int_{R}  x^{2} | (L(x,t)-1) F_{\sigma,\vec{{r}_0}}(x)| + \epsilon.
\end{eqnarray}
Recall that for finite $x$ interval, such as $R$, $(L(x,t)-1)$ uniformly goes to zero with vanishing $t$.  Hence, for any $\epsilon>0$ and $x \in R$ there exists a $t_{\epsilon}$ such that for all $|t| < t_{\epsilon}$ we have
\begin{eqnarray}
	\nu | K( t ) - 1 | & \leq & 2 \epsilon .
\end{eqnarray}
We conclude that $K( t )$ converges to 1 with vanishing $t$, proving our lemma.

We now employ the above lemmas to demonstrate pointwise convergence.  We wish to bound the quantity
\begin{equation}
	\delta(t, N) = | f_{\sigma,\vec{{r}_0}}(t/\sqrt{N})^{N} - e^{-\nu t^{2}/4} | ,
\end{equation}
which can be expanded as
\begin{eqnarray}
	\delta(t, N) & = & | (1 -  \frac{\nu  t^{2}}{4N}  K(t/\sqrt{N}) )^{N} - e^{-\nu t^{2}/4} | . 
\end{eqnarray}
In the limit of large $N$ we have that
\begin{eqnarray}
	\delta(t, N) & = & \left| \left(1 -  \frac{\nu t^{2}}{4N}  K(t/\sqrt{N}) \right)^{N} - \left(1- \frac{\nu t^{2}}{4N}  \right)^{N} \right| .
	\nonumber\\
\end{eqnarray}
For any $a$ and $b$ satisfying $|a|, |b| \leq 1$ it is well known that
\begin{equation} 
	| a^{N} - b^{N} | \leq N | a - b | .
\end{equation}
Applying this inequality to our problem gives
\begin{eqnarray}
	\delta(t, N) & \leq & N | (1 -  \frac{\nu t^{2}}{4N}  K(t/\sqrt{N}) ) - (1- \frac{\nu t^{2}}{4N} ) | ,\nonumber \\
	& \leq &   | \nu  t^{2}(  K(t/\sqrt{N}) - 1 )  | / 4.
\end{eqnarray}
For fixed $t$, the argument of $K(t / \sqrt{N})$ vanishes as $N$ increases, entailing pointwise convergence. Consider a point $t$, for any $\epsilon>0$ we can always find a $u_{\epsilon}$ such that $|K(u) - 1|\leq 4 \epsilon / (\nu t^{2})$ 
for all $|u| \leq u_{\epsilon}$. The variable $u_{\epsilon}$ defines a $N_{\epsilon}=t^{2}_{\epsilon}u_{\epsilon}^{2}$, such that $|u| =| t /\sqrt{N} | \leq u_{\epsilon}$ whenever $N>N_{\epsilon}$ and so
\begin{eqnarray}
	\delta(t, N>N_{\epsilon}) & \leq & \epsilon .
\end{eqnarray}
Hence we have shown Thm.~\ref{thm_QCLT}.  Furthermore,  this convergence is uniform on finite intervals of $t$.

\section{Convergence of expectation values}
\label{APP:expectationvalues}

In the letter we state that convergence of $\chi_{\sigma_{n}}$ to a Gaussian function with covariance matrix $\Gamma_{\sigma_{\infty}}$ entails convergence for the expectation value of operators, $B$, with absolutely integrable characteristic functions.   Here we prove the relevant lemma, and also show that its conditions are meet for operators of the form $B=\kb{x}{y}$ where $\ket{x}$ and $\ket{y}$ are,  up-to a Gaussian unitary, multimode Fock states.  Note that, when $\sigma_{n}$ are positive Hermitian operators, stronger conclusions can be reached by a similar means as shown in Ref.~\cite{CLT}.

\begin{lem}[Convergence of expectation values]
\label{lem_expectation}
	Consider a sequence of trace class m-mode operators $\sigma_{n}$ and a limiting operator $\sigma_{\infty}$, such that (i) $| \chi_{\sigma_{n, \infty}}(\vec{r}) | \leq 1$ for all $\vec{r}$ and (ii) For any compact region, $R$, $\chi_{\sigma_{n}}$ converges uniformly to $\chi_{\sigma_{\infty}}$.  Consider also a trace class operator $B$ satisfying $\int |\chi_{B}(\vec{r})| d\vec{r} < \infty$ and any $\epsilon>0$.  There exists an $n_{\epsilon}$ such that for all $n_{\epsilon} < n$
	\begin{equation}
		| \tr ( B \sigma_{n} ) - \tr ( B \sigma_{\infty} ) | \leq \epsilon .
	\end{equation}
\end{lem}
For the sequence of characteristic functions considered in the letter, property (i) is a prerequisite and property (ii) was shown in Cor.~(\ref{cor_uniform_converge}).  We state these properties again here for clarity.  First we note that for any two m-mode trace class operators, $B$ and $A$, the characteristic functions, $\chi_{B}$ and $\chi_{A}$, satisfy
\begin{equation} 
	  \int d\vec{r} \chi_{B} (\vec{r})  \chi_{A}(\vec{r})  = ( 2 \pi )^{m}  \tr ( B A ) .
\end{equation}
This entails that
\begin{eqnarray}
	D_{n} &= & (2 \pi)^{m}	| \tr ( B \sigma_{n} ) - \tr ( B \sigma_{\infty} ) | \\ &=&  | \int \chi_{B}(\vec{r}) [ \chi_{\sigma_{n}}(\vec{r}) - \chi_{\sigma_{\infty}}(\vec{r}) ] d \vec{r} | .
\end{eqnarray}
Next we observe that for any $\chi_{B}$ satisfying $\int |\chi_{B}(\vec{r})| d\vec{r} < \infty$, and for any $\epsilon > 0$ we can find a function $h$ with compact support such that
\begin{equation}
\label{Dense}
	\int | \chi_{B}(\vec{r}) -h(\vec{r}) | d\vec{r} \leq \epsilon .
\end{equation}
That is, the set of functions with compact support are dense in the set of absolutely integrable functions.  Defining $\eta(\vec{r}) = \chi_{B}(\vec{r}) -h(\vec{r}) $ and $\Delta_{n}(\vec{r})=\chi_{\sigma_{n}}(\vec{r})- \chi_{\sigma_{\infty}}(\vec{r})$, we have
\begin{eqnarray}
	D_{n} & = &    | \int [ \eta(\vec{r}) - h(\vec{r})] \Delta_{n} (\vec{r})d \vec{r} | , \\ \nonumber
	& \leq &  | \int  \eta(\vec{r})  \Delta_{n} (\vec{r}) d \vec{r} | +  | \int h(\vec{r})  \Delta_{n} (\vec{r}) d \vec{r} |  , \\ \nonumber
	& \leq &       \sup ( |\Delta_{n}(\vec{r}) |) \int | \eta(\vec{r}) |  d\vec{r}   +  | \int h(\vec{r})  \Delta_{n} d \vec{r} |   , \\ \nonumber
		& \leq & 2  \epsilon +  | \int h(\vec{r})  \Delta_{n} d \vec{r} |   .
\end{eqnarray}
The third line uses a H\"{o}lder's inequality.  The final line uses Eq.~(\ref{Dense}) and $|\Delta_{n}(\vec{r})| \leq 2$.  For sufficiently large $n$, the second term can also be made arbitrarily small because $h$ has compact support and $\Delta_{n}$ vanishes uniformly on any compact set, as required by property (ii).  This completes the proof of the lemma.

Furthermore, any operator $B=\kb{x}{y}$ will satisfy the conditions of the lemma when $\ket{x}$ and $\ket{y}$ are eigenvectors of a Gaussian density operator.  That is, there exists a Gaussian unitary such that $U\ket{x}$ and $U \ket{y}$ are Fock states.  Clearly $B$ is trace class, and so we only need show absolute convergence of $\chi_{B}$.  For a 1-mode element in the Fock basis $\kb{n}{m}$ the characteristic function will be a product of a real polynomial in $\vec{r}$ of finite degree and a decaying Gaussian in $|\vec{r}|$.  Such functions are easily seen to absolutely converge, and so too shall a product of $m$ such functions.  A Gaussian displacement only adds a phase factor to the characteristic function, and non-displacing Gaussian unitaries are equivalent to a unitary change of variables in the integral. Hence, Gaussian unitaries will not alter the value of the absolute integral.  Indeed, most trace class operators are easily seen to have absolutely integrable characteristic functions.  However, we do not currently know if this is generally true.

\section{Gaussian integrals and the Schur complement}
\label{APP:GaussANDSchur}

The product of two operators, e.g., $\rho \Pi$ has a characteristic function that equals an integral involving the characteristic functions for $\rho$ and $\Pi$ as expressed by Eq.~(\ref{eqn:fourier_convolve}) of the letter.  Furthermore, for Gaussian states this integral is solved by taking the Schur complement of an appropriate matrix (see Eq.~(\ref{eqn:Schur}) of the letter).  These techniques are standard, but for the purpose of completeness we shall review them here.

As a side remark, note that an alternative strategy towards formulating generalized
quantum central limit theorems would have been to consider objects of the form
$\Pi^{1/2} \rho \Pi^{1/2}/\tr(\Pi\rho)$ instead of $\Pi  \rho/\tr(\Pi\rho) $. While this constitutes
an alternative in principle, the resulting Schur complement expressions become much more 
involved, motivating the path taken above.

We begin by noting that for any trace class operator $A$, the formula for the characteristic function can be inverted such that
\begin{equation}
\label{eqn:invert_char}
	A \propto \int d \vec{r} \chi_{A}(\vec{r}) D(-\vec{r}).
\end{equation}
Hence, the product of two operators satisfies
\begin{equation}
	AB \propto \int \int d \vec{r} d \vec{r'} \chi_{A}(\vec{r}) 
	\chi_{B}(\vec{r'}) D(-\vec{r}) D(-\vec{r'}) .
\end{equation}
The Campbell-Baker-Haussdorf formula entails that
\begin{equation}
 	 D(-\vec{r}) D(-\vec{r'})  =  D(-(\vec{r}+\vec{r'})) \exp ( - i \vec{r}^{T} \Sigma \vec{r'} /2 ),
\end{equation}
and so
\begin{equation}
	AB \propto \int d \vec{r} d \vec{r'} \chi_{A}(\vec{r}) \chi_{B}(\vec{r'}) 
	D(-(\vec{r}+\vec{r'}))  \exp ( - i \vec{r}^{T} \Sigma \vec{r}' /2 ).
\end{equation}
Changing variables to $\vec{q}=\vec{r}+\vec{r'}$ and $\vec{r}=\vec{r}$ yields,
\begin{equation}
	AB \propto \int d \vec{r} d \vec{q} \chi_{A}(\vec{r}) \chi_{B}(\vec{q-r}) D(-\vec{q})  
	\exp ( - \frac{i}{2} \vec{r}^{T} \Sigma \vec{(q-r)} ).
\end{equation}
Noting that $\vec{r}^{T}\Sigma \vec{r}=0$ for all $\vec{r}$ eliminates part of the exponent.  Next we observe that this has the form of Eq.~(\ref{eqn:invert_char}) and so we can deduce that the characteristic function is
\begin{equation}
	\chi_{AB} (\vec{q}) \propto 
	\int d \vec{r} \chi_{A}(\vec{r}) \chi_{B}(\vec{q-r})  
	\exp ( - \frac{i}{2} \vec{r}^{T} \Sigma \vec{q} ).
\end{equation}
Setting $A=\rho_{\infty}$ and $B=\Pi$ provides Eq.~\ref{eqn:fourier_convolve} of the letter.  

Until now we have not used any properties of $A$ or $B$, but herein assume they are Gaussian with covariance matrices $\Gamma_{A}$ and $\Gamma_{B}$. Hence
\begin{eqnarray}
	\chi_{AB}(\vec{q}) & \propto & \int d \vec{r} 
	\exp \left( - \frac{\vec{r}^{T} \Gamma_{A}\vec{r}}{4}   \right) \\ 
	&\times &   \exp \left( - \frac{\vec{(q-r)}^{T} \Gamma_{B}\vec{(q-r)}}{4}    - 
	\frac{i}{2} \vec{r}^{T} \Sigma \vec{q} \right).\nonumber
\end{eqnarray}
This can be expressed more compactly using a single matrix
\begin{eqnarray}
	\chi_{AB}(\vec{q}) & \propto& \int \exp (- \vec{(\vec{q}, \vec{r})^{T}}M\vec{(\vec{q}, \vec{r})}/4  ) d\vec{r},
\end{eqnarray}
where $M$ is the Block matrix
\begin{equation}
	M  =  \left( \begin{array}{c c}  \Gamma_{B} & -\Gamma_{B} \\ 
	2 i \Sigma - \Gamma_{B} & \Gamma_{A}+\Gamma_{B}  \end{array} \right) .
\end{equation}
Observing that $\vec{r} \Sigma 
\vec{q}=-\vec{q} \Sigma \vec{r} $ allows explicit symmetrization,
\begin{equation}
	M  =  \left( \begin{array}{c c}  \Gamma_{B} & -\Gamma_{B}-i \Sigma \\ 
	i \Sigma - \Gamma_{B} & \Gamma_{A}+\Gamma_{B}  \end{array} \right) ,
\end{equation}
such that now $M=M^{T}$.  The Schur complements arises by decomposing the matrix as 
$M=N D N^{T}$, with
\begin{eqnarray}
	D  & = &  \left( \begin{array}{c c}  \mathrm{Sc}(M) & 0 \\ 0 & 
	\Gamma_{A}+\Gamma_{B}  \end{array} \right) ,\\
	N & = & \left(  \begin{array}{cc} \id 
	& -(\Gamma_{A}+i \Sigma )(\Gamma_{A}+\Gamma_{B})^{-1}  \\ 0 & \id  \end{array} \right) ,
\end{eqnarray}
where $\mathrm{Sc(M)}$ denotes the Schur complement of $M$,
\begin{equation}
	\mathrm{Sc(M)} = \Gamma_{B} - ( \Gamma_{B}+i \Sigma )
	( \Gamma_{A}+\Gamma_{B})^{-1}(\Gamma_{B}- i \Sigma ).
\end{equation}
This decomposition requires only that $M$ is symmetric and that $\Gamma_{A}+\Gamma_{B}$ is invertible.  Returning to our Gaussian integral, a change of variables 
$(\vec{q}, \vec{r})N^{T}\mapsto (\vec{q}, \vec{r})$ gives
\begin{eqnarray}
	\chi_{AB}(\vec{q}) & \propto& \int 
	\exp (- \vec{(\vec{q}, \vec{r})^{T}}D\vec{(\vec{q}, \vec{r})}/4  ) d\vec{r},
\end{eqnarray}
and so the $\vec{q}$ and $\vec{r}$ variables decouple, and the integral evaluates to some number and
\begin{eqnarray}
	\chi_{AB}(\vec{q}) & \propto& \exp (- \vec{q}^{T}\mathrm{Sc}(M) \vec{q}/4  ) d\vec{r}.
\end{eqnarray}
Hence, we have that $\Gamma_{AB} = \mathrm{Sc(M)}$.  By substituting in $\Gamma_{A}=\Gamma_{\rho_{\infty}}$ and $\Gamma_{B}=\Gamma_\Pi$ we obtain Eq.~(\ref{eqn:Schur}) of the letter.

\section{Notions of weak convergence} 
\label{APP:weaktostrong}

Here we briefly discuss technicalities related to notions of 
weak convergence versus convergence in trace-norm. Interestingly, 
these different notions can have a physical implication.  While in many experimentally relevant examples the conditions of Thm.~(\ref{thm2}) of the letter are satisfied, one can concoct exotic instances for which strong convergence fails.  For example, if we consider the $2$-mode state vector 
\begin{equation}
	\ket{\psi_{0}} \propto \ket{0,0}+0.1\ket{1,1}+6\ket{2,2} 
\end{equation}
and apply the GP protocol, then Thm.~(\ref{thm2}) of the letter holds for a target Gaussian 
\begin{equation}
	\ket{\psi_{\infty}} \propto \sum_{n} (1/10)^{n} \ket{n,n}. 
\end{equation}	
However, once correctly normalized, straightforward numerics show the state diverges with fidelity $\bk{\psi_{n}}{\psi_{\infty}}$ vanishing with $n$.  This highlights the importance of proving more robust convergence results. Stronger convergence has only been established for EP, where trivially $\tr(\rho_{n}\Pi)=\tr(\rho_{\infty}\Pi)$ since $\Pi=\id$.  On a theoretical level, it is compelling to ask what the necessary and sufficient conditions are for strong convergence of GG.   

\section{Iterative formula}
\label{AppIt}

Here we go through the details of the iterative formula of Eq.~(\ref{EqIt}).  For brevity, we assume $\Pi$ is diagonal in the Fock basis, and so $V=\id$, though the more general case can be proven by the same method but more cluttered notation.  For the $(n+1)th$ round we have that the state satisfies
\begin{eqnarray*}
	\sigma_{n+1}^{\vec{x},\vec{y}} & = & \frac{ \tr^{2} [ U( \rho_{n} \otimes \rho_{n}) U^{\dagger} ( \Pi \otimes \Pi )  ]}{  \tr [ U( \rho_{n} \otimes \rho_{n}) U^{\dagger} ( \Pi \otimes \Pi) ]  } \\ \nonumber
	 & = & \frac{ \tr^{2} [ U( \rho_{n} \otimes \rho_{n}) U^{\dagger} ( \Pi \otimes \Pi )  ]}{\tr [ \rho_{n} \Pi  ]^{2}}
\end{eqnarray*}
and so the expectation values satisfy
\begin{eqnarray*}
	\alpha_{n+1}^{\vec{x},\vec{y}} & = & \frac{ \tr [( \otimes_{k} \hat{a}_{k,1}^{x_{k}} )^{\dagger} ( \otimes_{j} \hat{a}_{j,1}^{y_{j}} ) U( \rho_{n} \otimes \rho_{n}) U^{\dagger} ( \Pi \otimes \Pi )  ]}{\tr [ \rho_{n} \Pi  ]^{2}}
\\ \nonumber
& = & \tr [U^{\dagger} ( \otimes_{k} \hat{a}_{k,1}^{x_{k}} )^{\dagger} ( \otimes_{j} \hat{a}_{j,1}^{y_{j}} ) U( \sigma_{n} \otimes \sigma_{n})  ] .
\end{eqnarray*}
Next we must find the effect of conjugating the unitary through the annihilation/creation operators.
\begin{equation*}
\begin{array}{lr}
	U^{\dagger}( \otimes_{k} \hat{a}_{k,1}^{x_{k}} )^{\dagger} ( \otimes_{j} \hat{a}_{j,1}^{y_{j}} ) U &   \\ 	
=  \left( \otimes_{k=1}^{m} \left(\frac{\hat{a}_{k,1}+\hat{a}_{k,2}}{\sqrt{2}} \right)^{x_{k}} \right)^{\dagger} \left( \otimes_{j=1}^{m} \left( \frac{ \hat{a}_{j,1}+\hat{a}_{j,2}}{\sqrt{2}} \right)^{y_{j}} \right) & \\	
=   \sum	  C_{\vec{u},\vec{v}}^{\vec{x},\vec{y}}  ( \otimes_{k}  \hat{a}_{k,1}^{u_{k}}  )( \otimes_{k}  \hat{a}_{k,2}^{x_{k}-u_{k}}  )   ( \otimes_{j}  \hat{a}_{j,2}^{v_{j}}  )( \otimes_{j}  \hat{a}_{j,2}^{y_{j}-v_{j}}  ) & 
	 \end{array}
\end{equation*}
where in the last line we have simply noted that all the binomial factors and contributions of $1/\sqrt{2}$ are non-negative and real numbers, and so sum to a non-negative number $ C_{\vec{u},\vec{v}}^{\vec{x},\vec{y}} $.   The exact value of these numbers are unimportant to us here.  We substitute this expression for the conjugated annihilation/creation operators into that for $\alpha_{n+1}^{\vec{x},\vec{y}}$.  Evaluating the trace, we replace the operators on the right hand side with the variables of the form $\alpha_{n}^{\vec{u},\vec{v}}$ and arrive at the desired formula.

\section{Trace norm convergence}
\label{TraceNorm}

Here we give our own version of the well known proof of trace norm convergence of $\rho_{n}$ to $\rho_{\infty}$ from point-wise convergence of the density matrix elements.  First we note  that for any $\epsilon>0$ there exists a projector $P$ of finite rank that commutes with $\Pi$, such that $\tr ( P \rho_{\infty}) > 1 - \epsilon$.  It is useful to denote $P= \sum_{x \in X}\kb{x}{x}$ where $X$ is a finite set of eigenvectors of $\Pi$.  Furthermore, we can find an $n_{\epsilon}$ such that for all $n> n_{\epsilon}$ we also have 
\begin{eqnarray}
	\tr( P\rho_{\infty} )  - 	\tr( P \rho_{n} )  < \epsilon .  
\end{eqnarray}
Combining this with $\tr ( P \rho_{\infty}) > 1 - \epsilon$, gives
\begin{eqnarray}
		\tr( P \rho_{n} )  > 1- 2\epsilon .  
\end{eqnarray}
Next we observe that the trace norm can be bounded as follows,
\begin{equation}
\label{F1}
\begin{array}{llr}
	&|| \rho_{n} - \rho_{\infty} ||_{1}  & \\ 	
	=& || (P ( \rho_{n} - \rho_{\infty}) P + ( \rho_{n} - P \rho_{n} P) - (\rho_{\infty}-P \rho_{\infty}P) ||_{1} & \\
	\leq& ||  P ( \rho_{n} - \rho_{\infty}) P ||_{1}  + || \rho_{n} - P \rho_{n} P ||_{1}  & \\ &+ ||\rho_{\infty}-P \rho_{\infty}P ||_{1} & \end{array}
\end{equation}
Let us consider the first term, which satisfies
\begin{equation}
	||  P ( \rho_{n} - \rho_{\infty}) P ||_{1} \leq \sum_{x, y \in  X} | \bra{y} \rho_{n} \ket{x} -\bra{y} \rho_{\infty} \ket{x} | .
\end{equation}
By our previously established results, for every pair $\ket{x}, \ket{y} \in X $, there exists an $n_{\epsilon, x, y}$ such that for all $n>n_{\epsilon, x, y}$, we have
\begin{equation}
	| \bra{y} \rho_{n} \ket{x} -\bra{y} \rho_{\infty} \ket{x} | < \epsilon / |X|^{2}
\end{equation}
Hence, for all $n>n*$ where $n*=\max \{n_{\epsilon, x, y} , n_{\epsilon} \}$ we have
\begin{equation}
	||  P ( \rho_{n} - \rho_{\infty}) P ||_{1} < \epsilon .
\end{equation}
We next consider the third term of Eq.~(\ref{F1}), noting that
\begin{eqnarray}
 	||\rho_{\infty}-P \rho_{\infty}P ||_{1} & \leq & 	2 || P \rho_{\infty}  (1-P) ||_{1}  , \\ \nonumber
	& & + || (1-P) \rho_{\infty}  (1-P) ||_{1} .
\end{eqnarray}
The later term is simply $||(1-P) \rho_{\infty}  (1-P)||_{1} = \tr ((1-P) \rho_{\infty}) < \epsilon $.  Since $\rho_{\infty}$ is Hermitian and positive we deduce
\begin{eqnarray*}
	|| P \rho_{\infty}  (1-P) ||_{1} & \leq & \sqrt{ || P \rho_{\infty} P ||_{1}   || (1-P) \rho_{\infty} (1-P) ||_{1}   } ,
\end{eqnarray*}
which easily shown to be an equality when $\rho_{\infty}$ is pure, with the inequality following for mixed states.  Noting $|| P \rho_{\infty} P ||_{1}  \leq 1$ and again $||(1-P) \rho_{\infty}  (1-P)||_{1} < \epsilon $ gives
\begin{eqnarray*}
	|| P \rho_{\infty}  (1-P) ||_{1} & < \sqrt{ \epsilon },
\end{eqnarray*}
and so
\begin{eqnarray}
 	||\rho_{\infty}-P \rho_{\infty}P ||_{1} & < &	2 \sqrt{\epsilon} + \epsilon .
\end{eqnarray}
By a similar argument we can also show that for all $n>n_{\epsilon}$,
\begin{equation}
	 || \rho_{n} - P \rho_{n} P ||_{1} <  \sqrt{2 \epsilon} + 2 \epsilon .
\end{equation}
Putting these pieces together entails that for all $n>n^{*}$,
\begin{equation}
	|| \rho_{n} - \rho_{\infty} ||_{1} <  4 \epsilon + (2+\sqrt{2}) \sqrt{\epsilon} .
\end{equation}
Hence, for any desired trace norm accuracy, and hence also fidelity, there always exists a number of iterations, $n*$, above which the protocol achieves this accuracy.

\end{document}